\documentclass[a4paper,11pt]{article}
\usepackage{jinstpub} 
\usepackage{lineno}

\proceeding{IPRD2023: 16$^{\text{th}}$ Topical Seminar on Innovative Particle and Radiation Detectors \\
25-29 Sep 2023\\
Siena (Italy)}

\title{\boldmath Plasma-Based Etching Approach for GEM Detector Microfabrication at FBK for X-ray polarimetry in space}

 \author[a,b,c,1]{A. Lega,\note{Corresponding author.}}
 \author[a]{D. Novel,}
 \author[a]{T. Facchinelli,}
 \author[d,e]{C. Sgrò,}
 \author[d,e]{L. Baldini,}
 \author[d,e]{M. Minuti,}
 \author[a]{M. Boscardin,}
 \author[a]{G. Pepponi,}
 \author[b,c]{R. Iuppa,}
 \author[a]{R. Hall-Wilton,}
 \author[d,e]{L. Latronico}

\affiliation[a]{Fondazione Bruno Kessler,\\Via Sommarive, 18, 38123 Povo, Trento, Italy}
\affiliation[b]{Università di Trento, Dipartimento di Fisica, \\Via Sommarive 14, 38123 Povo, Trento, Italy}
\affiliation[c]{Istituto Nazionale di Fisica Nucleare, Sezione di Trento, TIFPA, \\Via Sommarive, 14, 38123 Povo, Trento, Italy}
\affiliation[d]{Università di Pisa, Dipartimento di Fisica Enrico Fermi, \\Largo B. Pontecorvo 3, 56127 Pisa, Italy}
\affiliation[e]{Istituto Nazionale di Fisica Nucleare, Sezione di Pisa,\\Largo B. Pontecorvo 3, 56127 Pisa, Italy}

\emailAdd{alega@fbk.eu}

\abstract{Gas Electron Multiplier (GEM) detectors are crucial for enabling high-resolution X-ray polarization of astrophysical sources when coupled to custom pixel readout ASIC in Gas Pixel Detectors (GPD),  as in the Imaging X-ray Polarimetry Explorer (IXPE), the Polarlight cubesat pathfinder and the PFA telescope onboard the future large enhanced X-ray Timing and Polarimetry (eXTP) Chinese mission. The R\&D efforts of the IXPE collaboration have resulted in mature GPD technology. However, limitations in the classical wet-etch or laser-drilled fabrication process of GEMs motivated our exploration of alternative methods. This work focuses on investigating a plasma-based etching approach for fabricating GEM patterns at Fondazione Bruno Kessler (FBK). The objective is to improve the aspect ratio of the GEM holes, to mitigate the charging of the GEM dielectric which generates rate-dependent gain changes. Unlike the traditional wet-etch process, Reactive Ion Etching (RIE) enables more vertical etching profiles and thus better aspect ratios. Moreover, the RIE process promises to overcome non-uniformities in the GEM hole patterns which are believed to cause systemic effects in the azimuthal response of GPDs equipped with either laser-drilled or wet-etch GEMs. We present a GEM geometry with 20~$\mu$m in diameter and 50~$\mu$m pitch, accompanied by extensive characterization (SEM and PFIB) of the structural features and aspect ratios. The collaboration with INFN Pisa and Turin enabled us to compare the electrical properties of these detectors and test their performance in their use as electron multipliers in GPDs. Although this R\&D work is in its initial stages, it holds promise for enhancing the sensitivity of the IXPE mission in X-ray polarimetry measurements through GEM pattern with more vertical hole profiles. The outcomes of this study have the potential to advance the current technological platforms and improve the capabilities of future space-based X-ray polarimetry missions.}


\begin{document}
\maketitle
\flushbottom

\section{GEM in X-ray polarimetry astronomy}
Polarimetry, by revealing information about the magnetic fields, composition, and geometry of astronomical objects, has evolved into a fundamental observational technique in astrophysics. X-ray polarimetry in astronomy has seen remarkable growth in theoretical studies, instrument designs, and detector deployment on balloon experiments and missions. Various physical processes, including particle acceleration, emission in strong magnetic fields, and scattering on aspherical geometries, contribute to polarized X-rays in astrophysical sources. The studied sources range from galactic to extragalactic, featuring both point-like entities like black holes and neutron stars and extended structures like pulsar wind nebulae, supernova remnants, and molecular clouds, as described in \cite{Art:Ixpe_1, Art:Ixpe_2, Art:Ixpe_3}. Investigating X-ray polarization offers valuable insights into diverse astrophysical phenomena, making this field promising for advancing our understanding across different scales and environments. Crystal-based polarimeters, such as those utilizing calcite or quartz, have long been the backbone of polarimetric studies. These devices offer high precision and are effective across a broad range of wavelengths but the sensitivity is limited mainly allowing high-significance detections for a single bright source like the Crab Nebula \cite{Art:Crab}. Gas Pixel Detectors (GPD) \cite{Art:IXPE, Art:Bellazzini}, introduced in the early 2000s, revolutionized soft X-ray photoelectric polarimetry, increasing sensitivity by over an order of magnitude. The GPD is based on the integration of a Gaseous Electron Multiplier (GEM) with a specialized ASIC designed to quantify the photoelectron conversion products resulting from keV X-rays interacting with the detector gas. This innovative approach allows for the measurement of photon polarization by assessing the direction of the photoelectron generated during the conversion process within the gas medium \cite{Art:Ixpe_4}. This technological advancement enables the realization of mission concepts capable of observing multiple sources, thereby facilitating unprecedented and accurate polarimetric measurements. The primary technological hurdle associated with this detector lies in achieving precise alignment between the top and bottom metal hole arrays of the GEM. Given that the electron drift into the converting gas spans on the order of millimetres (using Dimethyl Ether, DME, in the IXPE context) and undergoes significant multiple scattering rather than following a straight path, the determination of its direction is primarily established within the initial hundreds of micrometres. Given that an individual hole in the IXPE GPD design measures around 30~$\mu$m, it is crucial to ensure a misalignment between the top and bottom metal layers that is below a few micrometres. Attaining such heightened precision poses a significant challenge, with a crucial bottleneck residing in the intricacies of the microfabrication process. The standard large-area GEM process, reliant on chemical etching through a PCB-like manufacturing approach, proves unfeasible for achieving the required precision. Advancing the precision of gas-based detectors necessitates pushing the limits of current microfabrication technologies. Investigating alternative techniques for the hole milling of GEMs provides an avenue to meet stringent precision standards. This study explores the feasibility of employing a plasma-based etching method for GEM microfabrication, leveraging the specialized silicon technologies laboratory at Fondazione Bruno Kessler (FBK). Unlike earlier studies, which predominantly focused on larger dimensions with holes ranging from 50 to 70~$\mu$m \cite{Art:Plasma}, we aim to fabricate smaller holes, specifically in the range of 20-30~$\mu$m or even less, as this aligns with the requirements of the IXPE project. Presently, IXPE utilizes laser-drilled GEMs \cite{Art:Laser,Art:Laser_2} with hole diameters of 30~$\mu$m, providing a dependable method for producing these small-sized features. As the project progresses, we aim to explore whether plasma etching techniques can yield improved results compared to the current laser-drilling approach.

\section{Plasma Focused Ion Beam GEM milling}

\begin{figure}
    \centering
    \includegraphics[scale=0.41]{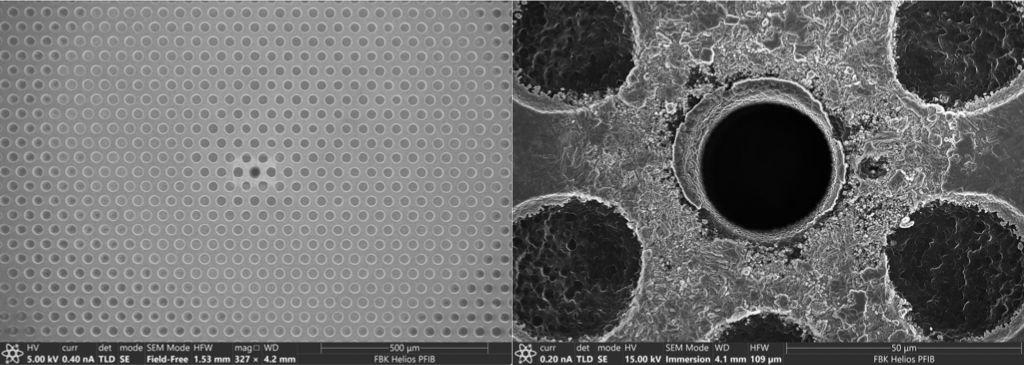}
    \caption{Patterned hole with an Xe-plasma ion beam from the PFIB. The hole dimensions are 30~$\mu$m in diameter and 50~$\mu$m in pitch, exhibiting a perfect cylindrical shape with no undercut. Due to the prolonged processing time of the PFIB, only one pattern was produced as a proof of concept. The copper (depicted in grey) has been affected by the plasma etching process.} 
    \label{fig:Img_2}
\end{figure}

\begin{figure}
    \centering
    \includegraphics[scale=0.28]{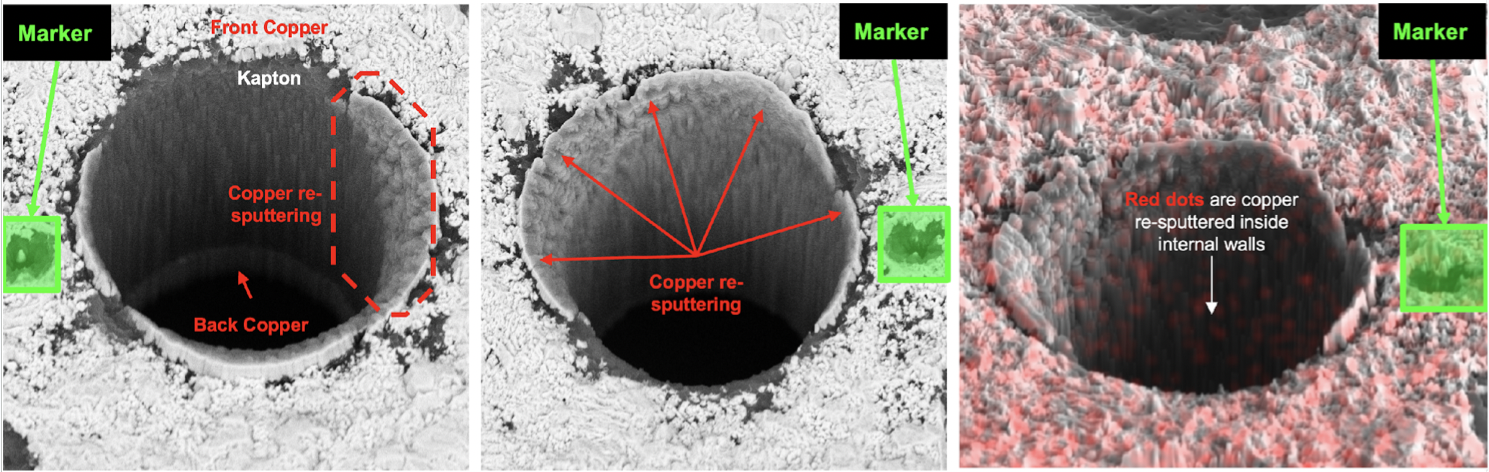}
    \caption{Energy Dispersive X-ray (EDX) analysis of the re-deposited copper.}
    \label{fig:Img_1}
\end{figure}

The first approach we propose in this work for small-size GEM hole milling is by using a Plasma Focused Ion Beam (PFIB) \cite{Art:PFIB}. The PFIB emerges as a sophisticated microfabrication tool, utilizing a Xenon ion beam generated from a plasma source. This technology finds applications in various fields, including semiconductor fabrication, materials science, and nanotechnology, primarily due to its ability to achieve high precision in material processing \cite{Art:PFIB_2}. The PFIB capability to attain sub-nanometer scale precision and precise control over the ion beam makes it an attractive technology for microfabrication processes, including the production of GEMs. The initial phase of the process entails patterning a thin substrate composed of copper (6~$\mu$m), poly-imide (50~$\mu$m), and copper (6~$\mu$m). In the second step, the definition of Cu holes occurs, involving the deposition of a photoresist layer onto the substrate. Photolithography techniques are then employed to expose the photoresist to UV light through a mask, thereby defining the desired pattern on the substrate. Subsequently, the exposed photoresist is developed, leaving behind a pattern that acts as a mask for the subsequent etching steps, carried out using standard FeCl$_3$ wet etching. PFIB facilitates direct-write micro-patterning, enabling the precise definition of patterns on the substrate using the focused ion beam. The ion beam is employed to selectively create the holes in the poly-imide substrate. Xe-plasma ions were used to microfabriacate a through hole via the patterned Cu-PI-Cu substrate. Additionally, PFIB systems commonly integrate in-situ imaging and monitoring capabilities, enabling real-time observation of the fabrication process. This feedback loop allows for on-the-fly adjustments during fabrication, ensuring the desired centring is maintained. The outcome of a singular hole milling using PFIB is illustrated in Fig. \ref{fig:Img_2}. During this proof-of-concept study, a single hole was milled with an hour-long patterning time per hole. Re-sputtering inside the hole can happen, thus requiring an extra cleaning step to achieve a vertical profile. This step limits the scale-up of the process automation with array processing. Given the extended processing timeline, engaging in full-pattern hole milling becomes impractical unless the milling strategy is optimized. Gas-assisted ion milling and higher ion currents could be implemented to speed-up processing times and avoid poly-imide resputtering. From the standpoint of undercut, it is negligible, showcasing a flawlessly cylindrical, state-of-the-art hole. It's noteworthy that the copper in the proximity of the milled hole has been sputtered away, resulting in a less clean pattern compared to its pre-milled state. This unintended feature is problematic, as copper may deposit into the holes, creating a conducting path between the two metals, which must be avoided to maintain the correct electric field and initiate the avalanche process. This effect is shown in Fig. \ref{fig:Img_1}, where the copper redeposited into the proximity of the hole is measured using Energy Dispersive X-ray (EDX) analysis. The confirmation of this behaviour was obtained with a simple electrical test, showing continuity between the top and bottom metals.

\section{Reactive Ion Etching GEM microfabrication}
\begin{figure}
    \centering
    \includegraphics[scale=0.33]{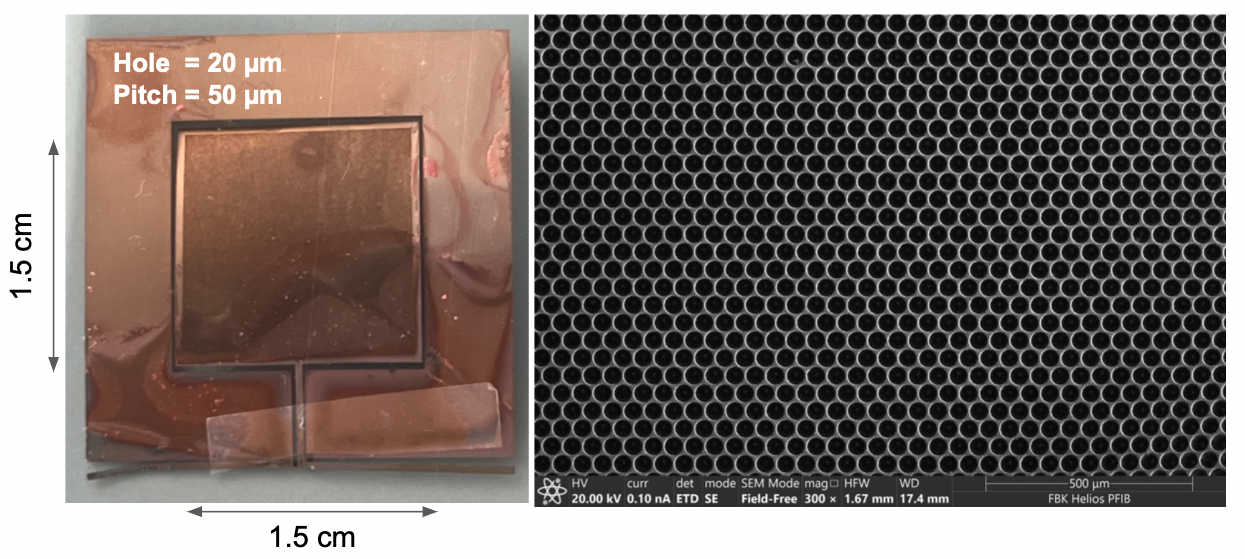}
    \caption{On the left side, the FBK RIE GEM sample is depicted, featuring its protective guard ring. On the right side, post-fabrication quality control imaging was conducted using PFIB in SEM mode.} 
    \label{fig:Img_3}
\end{figure}

The full-gem PFIB process seems impracticable, but the experience gained in the GEM plasma hole-milling for this first study allowed us to think of an alternative and more standard approach. The alternative methodology we advocate involves the application of Reactive Ion Etching (RIE). Good results for 50-70~$\mu$m standard holes are already reported in \cite{Art:Plasma}, but the technique was never implemented for the production of GEMs featuring small-sized holes (e.g. 20-30~$\mu$m). Reactive Ion Etching (RIE) technology plays a pivotal role in microfabrication processes, offering a highly controlled method for material removal and pattern definition \cite{Art:RIE}. When applied to the production of GEMs, RIE follows a well-defined sequence of steps to achieve precise and tailored results. RIE employs a reactive gas plasma to etch materials selectively depending on the substrate.
The initial steps of the process are identical to those of the PFIB, involving the patterning of holes in the metal layer. Following this, the substrate is subjected to Reactive Ion Etching (RIE) in the vacuum chamber. The reactive gas mixture, consisting of a combination of O$_2$/SF$_6$, is introduced into the RIE chamber where high-frequency voltage is applied to produce plasma. 
These reactive species chemically react with the material on the substrate, leading to its removal. The mask provided by the patterned Cu layer controls the areas of material that are etched and those that are protected. The reactive ions selectively etch the dielectric material, shaping the poly-imide holes with high accuracy. This step is crucial for the efficient multiplication of ionization signals within the GEM structure. RIE advantage lies in its ability to provide anisotropic etching, ensuring that the etching process is directional and well-defined. 
This characteristic facilitates the creation of high-aspect-ratio features with vertical sidewalls, enhancing the overall precision of the GEM structure. The aspect ratio achieved with the O$_2$ and SF$_6$ recipe is approximately 1:6, resulting in an undercut of about 8~$\mu$m in the 50~$\mu$m-thick poly-imide layer. Lithography and wet etching are then replicated on the backside of the GEM to achieve the back Cu patterning, with an alignment precision of about 2~$\mu$m. The patterned holes are illustrated in Fig. \ref{fig:Img_3}. Due to the limitations of the instrument chamber size (restricted to a 15~cm diameter circle, corresponding to the silicon 6" wafer standard.), this technique is suitable for producing only small-size GEMs. The chosen dimensions are 1.5x1.5~cm$^2$ to align with the IXPE ASIC for detector testing purposes. 

\section{Experimental characterization of RIE-GEM at INFN-Pisa}
\begin{figure}
    \centering
    \includegraphics[scale=0.4]{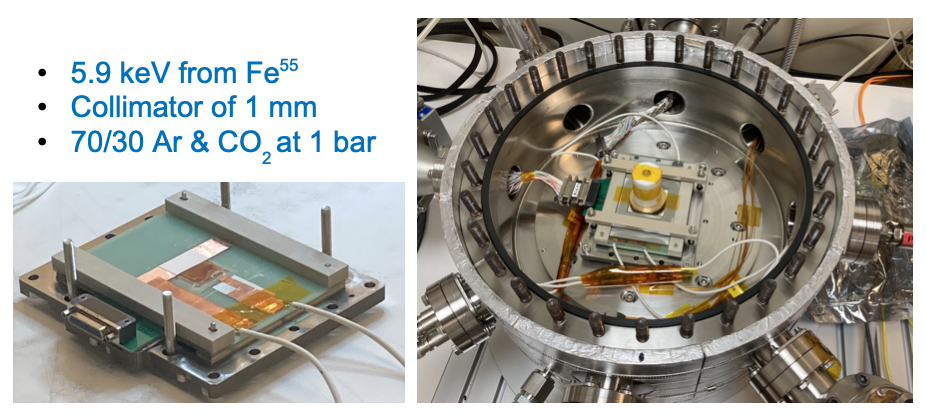}
    \caption{Testing of the FBK RIE GEM with a diameter of 20~$\mu$m and a pitch of 50~$\mu$m was conducted using a $^{55}$Fe radioactive source at INFN Pisa. On the left side, the assembled sample is integrated into the mounting mechanics, and on the right side, the vacuum setup chamber is shown, prepared to be filled with an Ar and CO$_2$ mixture at a ratio of 70/30.}
    \label{fig:Img_7}
\end{figure}

\begin{figure}
    \centering
    \includegraphics[scale=0.35]{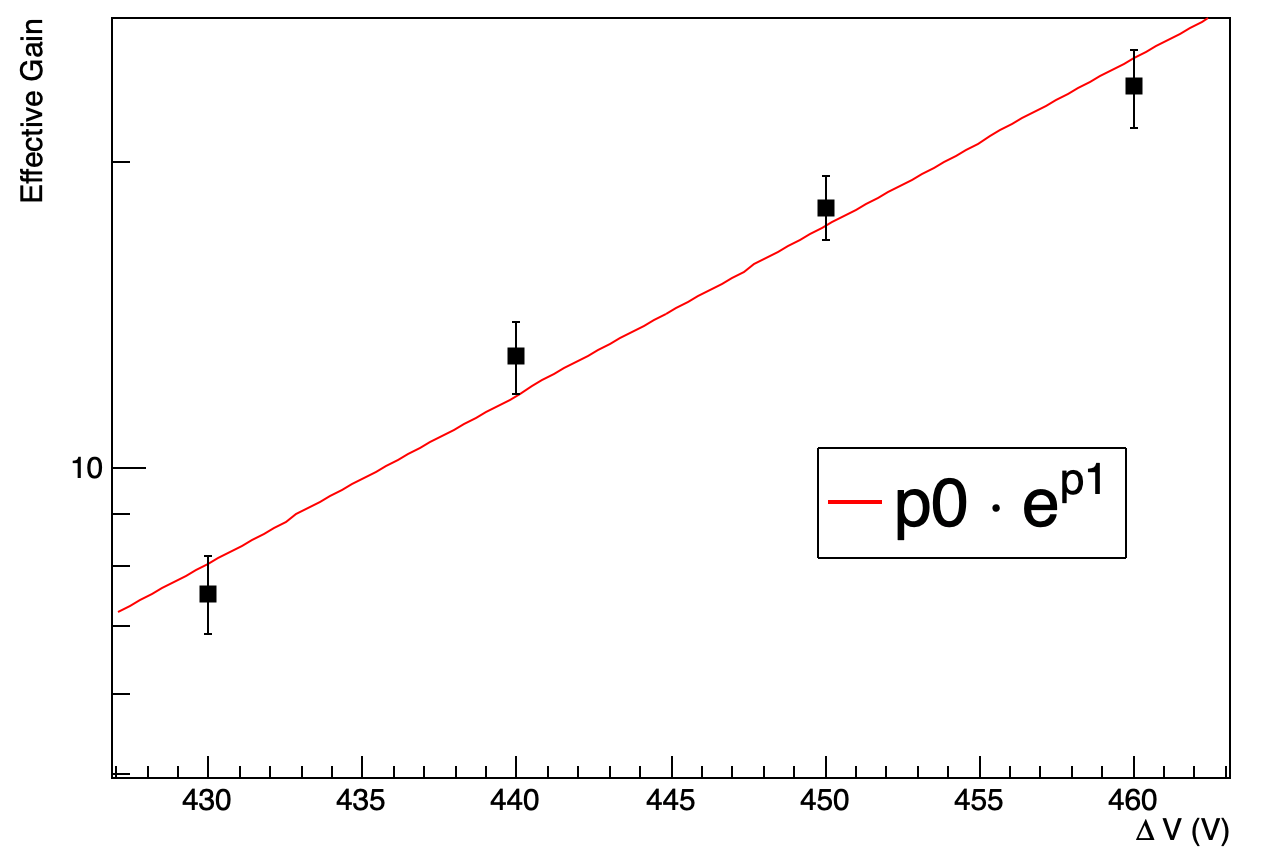}
    \caption{The effective gain, dependent on the applied voltage between the top and bottom layers of the GEM, is presented. Data points are selectively shown within the voltage range of 430~V (representing the initial gain point) to 460~V due to a discharge issue occurring between the top and bottom metals but outside the active area for voltages exceeding 460~V.} 
    \label{fig:Img_4}
\end{figure}

After the optimization of the microfabrication process, the first few samples underwent characterization through both a simple electrical test and exposure to a radioactive source within the desired energy range. The measurements and assessment of GEM performance took place at the INFN-Pisa laboratory. The characterization utilized the GEM-calibration setup developed for the IXPE mission, featuring an optimal diagnostic configuration for such experiments. To establish connections, the GEM was initially placed on a layer of conductive copper tape on both sides of the device. The fibreglass structure provided by the Pisa group served as support, and the GEM was secured using double-sided adhesive tape along the edges of the device (see Fig. \ref{fig:Img_7}, left). The GEM assembly took place in the vacuum setup of the Pisa group (see Fig. \ref{fig:Img_7}, right). The chamber was then filled with a mixture of 70\% argon and 30\% CO$_2$, maintaining a pressure of approximately 950 mbar. The $^{55}$Fe source, generating X-rays at 5.9~keV, was positioned within a collimation cylinder with a radius of approximately 1~mm. The initiation of electron amplification within the GEM was ascertained at a critical voltage differential, precisely at 430~V between the top and the bottom side. This observed initiation aligns seamlessly with anticipated outcomes derived from prior experiences involving GEMs of similar geometry within the context of IXPE \cite{Art:IXPE}. The device gain was validated by examining the cascades produced by X-rays exiting the collimated area on the ASIC. Attempts to increase the voltage between the top and bottom led to an electrical discharge at the GEM edges, particularly in the region where the bottom was in contact with the copper tape. This discharge exhibited Ohmic behaviour, generating a current of approximately 0.1-0.5 A. The discharge was likely caused by the proximity of the two metals at the edge, attributed to an approximately 8~$\mu$m undercut of the poly-imide. This region was the only one where pressure was applied to ensure good contact with the bottom, potentially causing the bending of the two metals in the protruding area. This issue arose because the same mask was used to define both sides (front and back) of the GEM active area. In the second part of the project, starting from the time of writing this article, a different approach which involves different masks for the patterning of the front and back will be employed. This is expected to address and solve the problem. Discharge-free events were analyzed to evaluate the gain. For four different voltages, the Pulse Height Amplitude ($PHA$), evaluated as the ADC measured by the ASIC, was recorded. Gaussian fits were applied to the distributions of $PHA$ for each voltage to extract the mean value. To calculate the effective gain, these values were multiplied by the number of electrons per ADC in the IXPE ASIC (as provided in \cite{Art:IXPE}) to convert the ADC readings into the corresponding number of electrons, employing the calibration factor:
\begin{equation}
    1~\text{ADC} = 3.7~e^-
\end{equation}
The minimum energy required to create an electron-ion pair in the mixture can be computed as:
\begin{equation}
    \text{E}_\text{thr}=25~eV
\end{equation}
Hence, the number of electron-ion pairs produced following the conversion of a photon into a photoelectron can be computed as:
\begin{equation}
    \text{Starting}_{e^-}=\frac{5900~eV}{25~eV}=236
\end{equation}
Thus, the final formula for evaluating the effective gain is:
\begin{equation}
        \text{Eff}_{\text{Gain}} = \frac{PHA_{\text{Mean}} \times 3.7}{236}
\end{equation}
The effective gain, as a function of the voltage applied between the top and bottom metal, is illustrated in Fig. \ref{fig:Img_4}. To enable a comparison with the gain curve presented in \cite{Art:IXPE} for IXPE, it is customary to fit the data with an exponential function to ascertain the voltage step required for a twofold gain increase. The result of the exponential fitting shows that the voltage increment needed to double the gain is about 18~V, a value similar to that reported for IXPE (around 25~V, \cite{Art:IXPE}). A comprehensive understanding of this result awaits an in-depth computational investigation utilizing a Monte Carlo approach, such as with Garfield++ \cite{Art:GIF}, which is scheduled to be undertaken in the upcoming months.

\section{Conclusions}
In this study, we explored alternative methods for the microfabrication of GEM detectors, focusing on a plasma-based etching approach at Fondazione Bruno Kessler (FBK). The primary goal was to surpass the constraints of conventional wet-etching or laser-drilling fabrication methods. The focus was on achieving enhanced aspect ratios for Gas Electron Multiplier holes and optimizing alignment between the two metal layers. The investigation included two innovative approaches: Plasma Focused Ion Beam (PFIB) GEM milling and Reactive Ion Etching (RIE) GEM microfabrication. PFIB demonstrated the potential for precise hole milling but faced challenges in scalability. On the other hand, RIE technology, a well-established microfabrication method, showcased promising results in creating GEM patterns with enhanced aspect ratios and reduced non-uniformities. Collaborating with INFN Pisa, we conducted extensive experimental characterizations of the RIE-GEM, assessing its electrical properties and performance in X-ray detection. Initial results indicated successful gain initiation at specific voltage configurations, with effective gains measured through a comprehensive analysis. However, challenges related to electrical discharges were identified, prompting adjustments in the fabrication approach. While this research is in its early stages, the outcomes hold promise for enhancing the sensitivity of X-ray polarimetry missions. The RIE-GEM fabrication process led to the creation of the first GEM microfabricated at FBK, signifying a breakthrough in the conventional silicon production focus of the centre. This proof of concept RIE-GEM fabrication represents the initial phase of a comprehensive study on GEMs of this type, particularly those featuring small-sized holes produced with plasma. The RIE-GEM fabrication process has the potential to advance current technological platforms, providing a foundation for future developments in space-based X-ray polarimetry missions. The study lays the groundwork for further investigations, including refining fabrication techniques, addressing electrical discharge issues, and conducting computational simulations to better understand the electric field dynamics within the detector. These efforts aim to contribute to the ongoing evolution of GEM detector technology, pushing the boundaries of microfabrication for improved performance in X-ray polarimetry applications and also towards the possibility of fabrication of smaller dimension features.



\begin{thebibliography}{99}

\bibitem{Art:Ixpe_1}
Weisskopf, M. C. et al., \textit{An imaging X-ray polarimeter for the study of galactic and extragalactic X-ray sources},  \href{https://www.researchgate.net/profile/Ronald-Elsner-2/publication/228712590_An_imaging_X-ray_polarimeter_for_the_study_of_galactic_and_extragalactic_X-ray_sources/links/00b7d5187d0118817d000000/An-imaging-X-ray-polarimeter-for-the-study-of-galactic-and-extragalactic-X-ray-sources.pdf}
  {Space Telescopes and Instrumentation 2008: Ultraviolet to Gamma Ray.ol. 7011. SPIE, 2008.}

\bibitem{Art:Ixpe_2}
Weisskopf, M. C.,  \textit{An overview of X-ray polarimetry of astronomical sources}, \href{https://www.mdpi.com/2075-4434/6/1/33}{Galaxies 6.1 (2018): 33.}

\bibitem{Art:Ixpe_3}
Deininger, W. et al., \textit{Small Satellite Platform Imaging X-Ray Polarimetry Explorer (IXPE) Mission Concept and Implementation, }\href{https://digitalcommons.usu.edu/cgi/viewcontent.cgi?article=4093&=&context=smallsat&=&sei-redir=1&referer=https%253A%252F%252Fscholar.google.com%252Fscholar%253Fhl%253Dit%2526as_sdt%253D0%25252C5%2526q%253DDeininger%25252C%252BW.%252Bet%252Bal.%25252C%252B%25255Ctextit%25257BSmall%252BSatellite%252BPlatform%252BImaging%252BX-Ray%252BPolarimetry%252BExplorer%252B%252528IXPE%252529%252BMission%252BConcept%252Band%252BImplementation.%25257D%252B%2525282018%252529.%2526btnG%253D#search=%22Deininger%2C%20W.%20et%20al.%2C%20%5Ctextit%7BSmall%20Satellite%20Platform%20Imaging%20X-Ray%20Polarimetry%20Explorer%20%28IXPE%29%20Mission%20Concept%20Implementation.%7D%20%282018%29.%22}{ 2018}.

\bibitem{Art:Crab}
Weisskopf, M. C. et al., \textit{A precision measurement of the X-ray polarization of the Crab Nebula without pulsar contamination, }\href{https://adsabs.harvard.edu/pdf/1978ApJ...220L.117W}{Astrophysical Journal, Part 2-Letters to the Editor,ol. 220, Mar. 15, 1978, p. L117-L121. 220 (1978): L117-L121.} 

\bibitem{Art:IXPE}
Baldini, L. et al.,
\textit{Design, construction, and test of the Gas Pixel Detectors for the IXPE mission},\href{https://www.sciencedirect.com/science/article/pii/S0927650521000670}{
Astroparticle Physics133. 102628 (2021).}


\bibitem{Art:Bellazzini}
Bellazzini, R. et al., \textit{Reading a GEM with aLSI pixel ASIC used as a direct charge collecting anode, } \href{https://www.sciencedirect.com/science/article/pii/S0168900204017103?casa_token=U3tZehX3Zz0AAAAA:-Cpjxy68HfUAasTY_3C5RzqZhbD0_vuX7DBBBgTQT_cb3JfUXpboNmDE2x9V40K4rqirWkq6mQ}{Nuclear Instruments and Methods in Physics Research Section A: Accelerators, Spectrometers, Detectors and Associated Equipment 535.1-2 (2004): 477-484.}

\bibitem{Art:Ixpe_4}
Costa, E. et al., \textit{An efficient photoelectric X-ray polarimeter for the study of black holes and neutron stars, } \href{https://www.nature.com/articles/35079508}{Nature 411.6838 (2001): 662-665.}

\bibitem{Art:Plasma}
Inuzuka, M. et al., \textit{Gas electron multiplier produced with the plasma etching method, }\href{https://www.sciencedirect.com/science/article/pii/S0168900204002244?casa_token=alnNghnIXNQAAAAA:W4RPVNt31nyNOQkWiFgFJ199LU0M1TLn7X-0xukuPvt4ULaKu6iXWVJMRWMcWsCROYB5LImZnQ}{Nuclear Instruments and Methods in Physics Research Section A: Accelerators, Spectrometers, Detectors and Associated Equipment 525.3 (2004): 529-534.}

\bibitem{Art:Laser}
Wu, X. et al., \textit{Development of a Novel Fabrication Process for Application in Glass Gas Electron Multiplier Detectors, }\href{https://www.mdpi.com/2227-9717/11/4/1215}{Processes 11.4 (2023): 1215.}

\bibitem{Art:Laser_2}
Tamagawa, T. et al., \textit{Development of gas electron multiplier foils with a laser etching technique, }\href{https://www.sciencedirect.com/science/article/pii/S0168900206000672?casa_token=XF9gNv8r-8QAAAAA:-hMjBEj7El7is8K5z7hpsk0qc0DW4JtMSk5TLU9adK8CXFEpnJydVD0X9DJCyvJeTcxVb4yRHQ}{Nuclear Instruments and Methods in Physics Research Section A: Accelerators, Spectrometers, Detectors and Associated Equipment 560.2 (2006): 418-424.}

\bibitem{Art:PFIB}
Burnett, T. L. et al, \textit{Largeolume serial section tomography by Xe Plasma FIB dual beam microscopy, } \href{https://www.sciencedirect.com/science/article/pii/S0304399115300644}{Ultramicroscopy 161 (2016): 119-129.}

\bibitem{Art:PFIB_2}
Utlaut, M. \textit{Focused ion beams for nano-machining and imaging, }\href{https://www.sciencedirect.com/science/article/abs/pii/B978085709500850004X}{Nanolithography. Woodhead Publishing, 2014. 116-157.}

\bibitem{Art:RIE}
Huff, M. \textit{Recent advances in reactive ion etching and applications of high-aspect-ratio microfabrication, } \href{https://www.mdpi.com/2072-666X/12/8/991}{Micromachines 12.8 (2021): 991.}

\bibitem{Art:GIF}
Veenhof, R. \textit{Garfield, a drift chamber simulation program.} \href{https://www.worldscientific.com/doi/abs/10.1142/9789814534598}{Conf. Proc. C.ol. 9306149. 1993.}

\end{thebibliography}


\acknowledgments
We express our gratitude to Fondazione Bruno Kessler (FBK) for their essential resources and support, and INFN Pisa and Turin for their collaboration in RIE-GEM design and characterization. This study was conducted within the framework of the FBK and INFN MiNaTAP agreement.

\bibliographystyle{JHEP}

\end{document}